\documentclass[aps,prl,twocolumn,superscriptaddress,showpacs]{revtex4}
\usepackage{graphicx}

\usepackage{dcolumn}

\usepackage{bm}
\usepackage{color}
\usepackage[colorlinks]{hyperref}

\begin{document}


\title{Radiation-induced oscillatory-magnetoresistance in a tilted magnetic field in $GaAs/Al_{x}Ga_{1-x}As$ devices}

\author{R. G. Mani}
\affiliation{Harvard University, Gordon McKay Laboratory of
Applied Science, 9 Oxford Street, Cambridge, MA 02138 U.S.A}

%
%
%
%
\date{\today}
\begin{abstract}
We examine the microwave-photoexcited magnetoresistance
oscillations in a tilted magnetic field in the high mobility
two-dimensional electron system. In analogy to the 2D Shubnikov-de
Haas effect, the characteristic field, $B_{f}$, and the period of
the radiation-induced magnetoresistance oscillations appears
dependent upon the component of the applied magnetic field that is
perpendicular to the plane of the 2DES. In addition, we find that
a parallel component, $B_{//}$, in the range of $0.6 < B_{//} <
1.2$ Tesla, at a tilt angle of $\theta = 80^{0}$, leaves the
oscillatory pattern essentially unchanged.
\end{abstract}
%
\pacs{73.21.-b,73.40.-c,73.43.-f; \textit{Journal Ref}: Phys. Rev. B\textbf{72}, 075327 (2005)}
%
\maketitle
\section{introduction}
The possibility of inducing unusual zero-resistance states and
magnetoresistance oscillations by photo-exciting a high mobility
GaAs/AlGaAs device, with radiation from the microwave and
Terahertz parts of the electromagnetic wave
spectrum,\cite{1,2,3,4,5,6} has recently motivated a broad
theoretical examination of the photoexcited steady states of the
low dimensional electron
system.\cite{7,8,9,10,11,12,13,14,15,16,17,18}

At the present, the observed radiation-induced resistance
oscillations are generally attributed to a field dependent
scattering at impurities and/or a steady state change in the
electronic distribution function, as a result of
photoexcitation.\cite{7,9,11,12,13,14,15} It turns out that in
both of these theoretical scenarios, the amplitude of the
magnetoresistance oscillations can increase with the radiation
intensity, in analogy to the experimental observations.
Consequently, the calculated resistivity or conductivity can be
made take on negative values at the minima of the oscillatory
magnetoresistivity (or magnetoconductivity) for sufficiently large
radiation intensities.\cite{7,9,11,12,14,15} A path for realizing
zero-resistance states from the theoretically-indicated negative
resistivity/conductivity under photoexcitation has been provided
by Andreev et al.,\cite{10} who suggested that a physical
instability of the negative resistivity/conductivity state should
transform it into a zero-resistance state, through the development
of dissipation-less current domains. In this approach, the current
domains reconfigure themselves to accommodate changes in the
applied current.\cite{10}

Recently, an alternate scenario has been provided by Inarrea and
Platero,\cite{17} who suggest that  a blocking of the final states
for scattering, due to an exclusion principle, leads to the
zero-resistance states observed in experiment. Although there
exist other models, the above mentioned theories seem to
constitute the popular approaches for understanding the observed
phenomena.

Transport studies in a tilted magnetic field have been utilized in
the past to establish the effective system-dimensionality in
electronic transport. In quasi 2-Dimensional Electronic Systems
(2DES), they have also served to separate the relative
contributions of spin and orbital effects. For example, it is
known that in an applied magnetic field $B$, when the 2DES
specimen is tilted at an angle $\theta$, the period of Landau
quantization dependent (orbital) effects, such as the Shubnikov-de
Haas (SdH) effect, is determined by the sample-perpendicular
magnetic field component $B_{\perp}$. On the other hand, it is
also known that spin related phenomena typically depend upon $B$
instead of $B_{\perp}$ since the spin degree of freedom couples to
the total applied magnetic field.\cite{19}

We examine here the radiation-induced oscillatory resistance in a
tilted magnetic field to experimentally confirm the effective
dimensionality, and examine the relative influence of the
perpendicular and in-plane ($B_{//}$) components of the applied
magnetic field.\cite{1} This extended report seems timely in light
of the recent observation of an anomalous disappearance of
radiation-induced zero-resistance states and associated
magnetoresistance oscillations under the application of a small (
$B_{//} \approx 0.5$ Tesla) parallel magnetic field on the 2D
electron system.\cite{20} A summary of our tilt field studies
appeared in Ref. 1.
\begin{figure}[b]
\begin{center}
\leavevmode
\includegraphics[scale = 0.25,angle=0,keepaspectratio=true,width=2.75in]{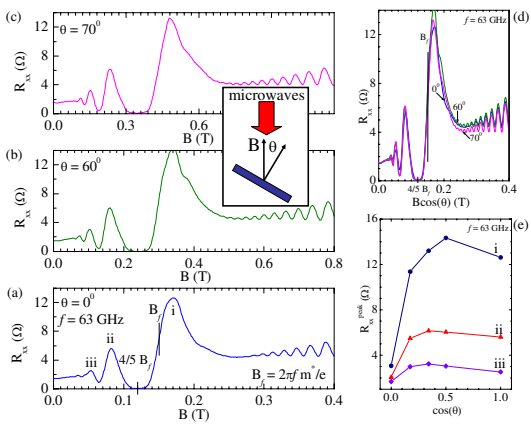}
\end{center}
\caption{ This figure illustrates the effect of tilting the
microwave excited two-dimensional electron system with respect to
the magnetic field and the direction of microwave propagation. The
tilt angle $\theta =$ (a) $0^{0}$, (b) $60^{0}$, and (c) $70^{0}$.
Note that the $B$-field scale shown on the abscissa of each panel
increases with increasing tilt angle $\theta$. (d) The data of (a)
- (c) have been plotted vs. $B \cos (\theta)$ in order to show
that the characteristic field $B_{f}$ and period of the radiation
induced oscillatory magnetoresistance is determined by the
perpendicular component, $B_{\perp}$ = $B\cos (\theta)$ of applied
magnetic field $B$. (e) The values of the resistance $R_{xx}$ at
the peaks in plot (a), labelled (i), (ii), and (iii), have been
plotted vs. $\cos (\theta)$.} \label{mani01fig}
\end{figure}
Briefly, we find that the characteristic field scale, $B_{f}$, or
equivalently the periodicity ($B_{f}^{-1})$, of the observed
oscillations is determined by $B_{\perp}$, analogous to the
characteristics of the 2D Shubnikov-de Haas effect.\cite{21} In
addition, the applied $B_{//}$ seems not to quench the observed
phenomena to a tilt angle $\theta$ = 80$^0$, although the
radiation-induced magnetoresistance oscillations and associated
zero-resistance states do disappear in the $\theta$ $\rightarrow$
$90^{0}$ limit.

\section{experiment}

For these experiments, specimens characterized by $n \approx 3
\times 10^{11} cm^{-2}$ and $\mu \leq 1.5 \times 10^{7} cm^{2}/Vs$
were mounted within a microwave waveguide, and immersed in liquid
Helium in a low temperature cryostat, within the bore of a
superconducting solenoid.\cite{1} \emph{In situ} sample rotation
was carried out by fixing the sample on a geared rotatable
platform, which could be turned with the aid of a geared shaft
that extended outside the cryostat. The quoted tilt angles,
$\theta$, are the mechanically set values, which could be
incremented in units of $10^{0}$. The gear backlash and/or play
produces an uncertainty in the orientation of up to $\pm 2 ^{0}$.
Thus, the actual tilt angle can differ slightly from the preset
value. This difference becomes consequential especially for
$\theta = 90^0$ case, which has therefore been denoted as $\theta
\approx 90^0$. An estimate of the effective tilt angle,
$\theta_{eff}$, was obtained from the data analysis, and these are
also indicated at the appropriate point in the discussion. The
electrical measurements were carried out using standard low
frequency \textit{ac} lock-in techniques in an oversized condition
for the waveguide (at 63 GHz), which implies an indeterminate
microwave polarization. The microwave intensity was preset to an
optimal value and remained undisturbed throughout the experiment.

\section{results}

Fig. 1(a) shows the low-$B$ transport under photoexcitation at 63
GHz with the specimen oriented at zero tilt-angle, i.e., $\theta$
= 0, (see inset Fig. 1), such that the sample normal lies parallel
to the applied magnetic field, $B$. In all these measurements, the
axis of propagation of the electromagnetic waves lies parallel to
the magnetic field axis, and the electric field of the microwaves
lies in the plane perpendicular to $B$. Fig. 1(a) indicates a wide
radiation-induced zero-resistance state in $R_{xx}$ about
(4/5)$B_{f}$, and a close approach to vanishing resistance at the
next lower-$B$ minimum, near (4/9)$B_{f}$, which follow the series
$B = (4/[4j+1])B_{f}$, with $j$ = 1,2,3.... Here, $B_{f}$ = $2\pi
f m^{*}/e$, $m^{*}$ is an effective mass, $e$ is the electron
charge, and $f$ is the radiation frequency.\cite{1}

The effect of tilting the specimen with respect to the magnetic
field is illustrated in Figs. 1(b) and 1(c). These figures
indicate self-similarity in the oscillatory resistance pattern
under tilt, provided that the magnetic field scale is increased
appropriately with increasing tilt angle. Such data suggest that
the characteristics field scale, $B_{f}$, in the absence of tilt,
i.e., $\theta = 0$, is mapped onto $B_{f}/\cos(\theta)$ at a
finite tilt angle, $\theta$, reflecting a dependence of the
underlying phenomena on $B_{\perp}$ = $B\cos(\theta)$ [see Fig.
1(d)]. This plot, Fig. 1(d), exhibits data collapse and confirms
that $B_{\perp}$ sets the characteristic field $B_{f}$ and the
inverse-magnetic-field periodicity $B_{f}^{-1}$ of the
radiation-induced resistance oscillations. The analysis indicated,
in addition, a small difference between the preset tilt angle,
$\theta$, and the effective tilt angle, $\theta_{eff}$, in the
data, which is attributed here to an orientational uncertainty.
Thus, we report that for $\theta = 60^0$, $\theta_{eff} = 60.2^0$,
and for $\theta = 70^0$, $\theta_{eff} = 69.7^0$.

Although the oscillatory resistance traces show similarity under
tilt when plotted versus $B_{\perp}$, there do occur some
systematic variations in the data from one tilt angle to another.
For example, a close comparison of Figs. 1(a) - (c) indicates a
non-monotonic variation in the amplitude of the radiation induced
oscillations with increasing tilt angle [see Fig. 1(e)]. In
particular, the resistance peak that has been labelled as (i) in
Figs. 1(a) and 1(e), increases in height in going from Fig. 1(a)
to 1(b), and then decreases in height from Fig. 1(b) to 1(c). We
explain this effect as follows:  It turns out that, in experiment,
the oscillation peak amplitude initially increases, then
saturates, and finally decreases with increasing radiation
intensity.\cite{1} In the experimental data shown in Fig. 1, the
radiation intensity at $\theta = 0^{0}$ corresponds to an
overexcited condition, a regime where the amplitude tends to
decrease with increasing intensity. Such an over-excited condition
was chosen for the $\theta = 0^{0}$ measurement in order to
realize a significant photon flux on the 2DES even at the highest
tilt angles. Thus, as the tilt angle is increased from $\theta =
0^{0}$ to $\theta = 60^{0}$, the effective photon flux on the
sample decreases, but this leads to a counter-intuitive increase
in the peak amplitude with increasing tilt angle [Fig. 1(a) and
1(b)]. Upon moving to the higher tilt angle, $\theta = 70^{0}$ in
Fig. 1(c), the oscillation peak amplitude (i) now decreases
because this corresponds to the regime where a decrease in
excitation intensity produces also a decrease in the oscillatory
resistance amplitude.

The effect of increasing the tilt angle to nearly $90^{0}$ is
illustrated in Fig. 2, where we have compared the data traces
obtained at $\theta = 0^{0}$ and $\theta \approx 90^{0}$, with the
$\theta \approx 90^{0}$ data extending to 11.5 Tesla, close to the
maximum rated magnetic field of the superconducting magnet. Here,
in Fig. 2(b), the weak oscillations in the vicinity of $B = 4$
Tesla and $B = 8$ Tesla are indicative of the virtual
disappearance of the radiation-induced resistance oscillations and
associated zero-resistance state in the $\theta \rightarrow
90^{0}$ limit. Note that in this situation, essentially all of the
applied magnetic field, $B$, appears as a component, $B_{//}$, in
the plane of the 2DES. Here, the effective tilt angle extracted
from the data is $\theta_{eff} = 88.55^0$.
\begin{figure}[t]
\begin{center}
 \leavevmode
\includegraphics[scale = 0.25,angle=0,keepaspectratio=true,width=2.75in]{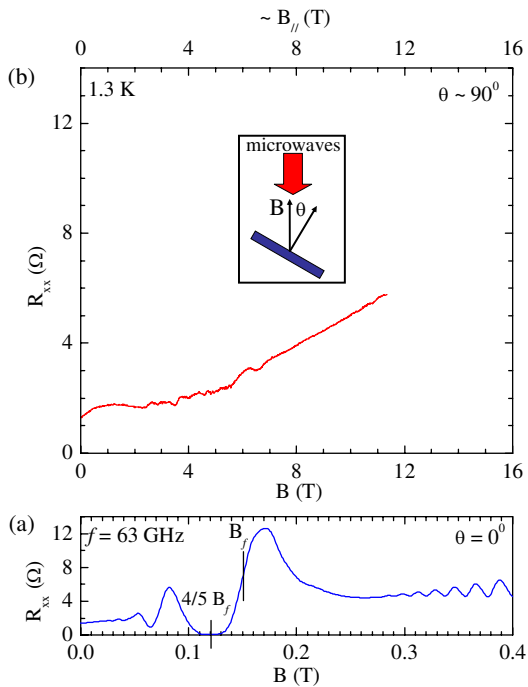}
\end{center}
\caption{ This figure compares the oscillatory photoinduced
magnetoresistance characteristics observed at a tilt angle $\theta
= 0^{0}$ (a) with the data obtained at $\theta \approx 90^{0}$
(b). Note the enormously enhanced field scale for the $\theta
\approx 90^{0}$ case. These data suggest that the
radiation-induced resistance oscillations vanish in the $\theta
\rightarrow 90^{0}$ limit.} \label{mani02fig}
\end{figure}

At the outset, one tends to attribute such a reduction in the
amplitude of the radiation-induced resistance oscillations in the
large tilt-angle limit to the vanishing photon flux on the 2DES,
when the 2DES is oriented nearly parallel to the direction of
microwave propagation. Yet, from the data of Fig. 2(b), it is
difficult to rule out the alternate possibility that it is the
application of a large parallel magnetic field, $B_{//}
> 3$ Tesla, which plays some unforseen role in the quenching of
the oscillations.

To address this point, we compare in Fig. 3, the data traces
obtained at $\theta = 0^{0}$ and $\theta = 80^{0}$. At $\theta =
80^{0}$, where the data suggest $\theta_{eff} = 79^0$, the
$B_{//}$ provided at the top abscissa of Fig. 3(b) indicates that
a large fraction of the applied $B$ appears as an in-plane
component. Yet, the radiation-induced resistance oscillations
continue to be observable at $\theta = 80^{0}$, with just a small
reduction in the peak height in comparison to the $\theta = 0^{0}$
condition. From these data, it seems possible to conclude that,
for the highest radiation-induced $R_{xx}$ peak and the deepest
$R_{xx}$ valley, a parallel magnetic field component that lies
between $0.6 < B_{//} < 1.2$ Tesla fails to extinguish the typical
characteristics. Thus, such data support the hypothesis that the
vanishing photon flux is the cause for the disappearance of the
radiation-induced oscillatory resistance in the $\theta
\rightarrow 90^{0}$ limit in our tilted magnetic field
experiments. In comparison, the amplitude of the oscillations in
the 2D SdH effect is not expected to vanish in a similar $\theta
\rightarrow 90^{0}$ limit (see Ref. 19 and, for example, Fig. 1(d)
and 3(b), right inset), although the $B$ required to realize the
associated oscillations becomes tremendously large, in the large
tilt angle limit.

\section{discussion}

Here, we relate our observations to the recent report of a strong
suppression of the radiation-induced zero-resistance states and
associated magnetoresistance oscillations by a small parallel
magnetic field, $B_{//}$.\cite{20} As evident from Figs. 1-3, and
especially Fig. 3, our data do not confirm that a modest $B_{//}$,
$B_{//} \leq 0.5$ T, causes a strong suppression of the
radiation-induced zero-resistance states and associated
magneto-resistance oscillations. Although there are experimental
differences between their two-axes magnet measurements\cite{20}
and our tilt-field measurements, such differences seem unlikely to
be the cause for the observed discrepancy. Perhaps, the
dissimilarities in observations are rooted in subtle differences
in the physical environments between our respective specimens. For
example, in the specimens utilized in ref. 20, the onset of SdH
oscillations moves to higher magnetic fields with increasing
$B_{//}$. In comparison, the SdH data exhibited in the right inset
of Fig. 3(b) seem not to indicate a similar shift of the SdH onset
to higher $B_{\perp}$ with increasing $\theta$. The shift of the
SdH onset to higher perpendicular $B$ in ref. 20 could be a
signature of an increase in the Landau level broadening in the
presence of a parallel magnetic field. Plausibly, such a change in
level broadening could then produce a chain reaction, including
the observed disappearance of the radiation-induced
zero-resistance states and associated resistance
oscillations.\cite{20} Yet, in such a scenario, it is not clear
why a fifty percent increase in the field for the onset of SdH
oscillations leads to the complete disappearance of the
radiation-induced effects in Ref. 20.

It could also be that small differences in the physical
environment become especially significant in the highest mobility
specimens, and that our lower mobility specimens provide for some
stabilization against such perturbations. Further experimental
studies appear necessary, however, to obtain further understanding
of the observed discrepancy. At the moment, it appears that the
parallel magnetic field component produces a variable response in
the photoexcited 2DES.

\begin{figure}[t]
\begin{center}
\leavevmode
\includegraphics[scale = 0.25,angle=0,keepaspectratio=true,width=2.75in]{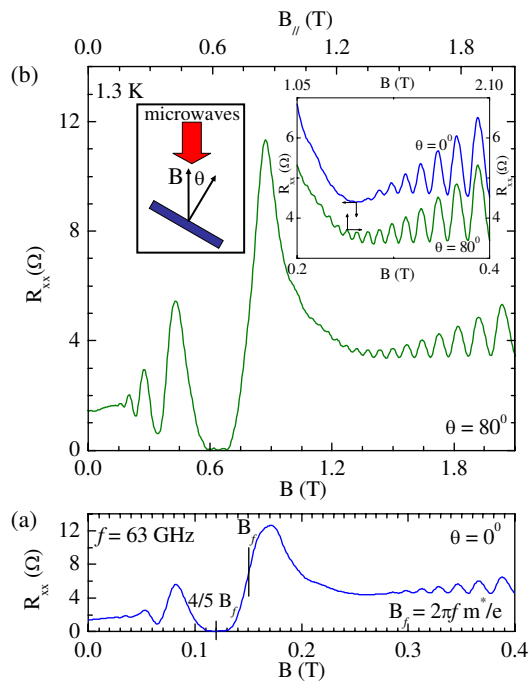}
\end{center}
 \caption{ This figure compares the characteristics observed at a
tilt angle $\theta = 0^{0}$ (a) with the data obtained at $\theta
= 80^{0}$ (b). In (b), the component of the applied magnetic field
that is parallel to the 2DES is indicated as $B_{//}$ [$= B
sin(\theta)]$ along the top abscissa. Note the similarity in the
oscillatory traces at $\theta = 0^{0}$ and $\theta = 80^{0}$. The
right inset to (b) compares the SdH oscillations at $\theta =
0^{0}$ and $80^{0}$.} \label{mani03fig}\end{figure}

\section{summary}

In summary, we have examined the effect of tilting a photoexcited
high mobility 2DES with respect to the applied magnetic field. We
find that the characteristic field $B_{f}$ and the
inverse-magnetic-field periodicity of the radiation-induced
magnetoresistance oscillations is determined by the component of
the magnetic field that is perpendicular to the 2DES [Fig. 1(d)],
as is typical for a 2D orbital effect. In addition, a parallel
magnetic field component, $B_{//}$ in the range $0.6 < B < 1.2$ T
at a tilt angle $\theta = 80 ^{0}$ appears insufficient to quench
the observed photoinduced oscillatory magnetoresistance. Yet, the
radiation induced resistance oscillation do disappear in the
$\theta \rightarrow 90^{0}$ limit, as the effective photon flux on
the 2DES vanishes.

\section{acknowledgements}

We acknowledge discussions with K. von Klitzing, V. Narayanamurti,
J. Smet, and W. Johnson. The high quality MBE material was
expertly grown by V. Umansky.

  \vspace{0cm}


\begin{thebibliography}{21}
\section{references}

\bibitem{1} R. G. Mani, J. H. Smet, K. von Klitzing, V. Narayanamurti, W. B.
Johnson, and V. Umansky, Nature (London) \textbf{420}, 646 (2002);
Phys. Rev. B \textbf{69}, 193304 (2004); Phys. Rev. Lett.
\textbf{92}, 146801 (2004); R. G. Mani, V. Narayanamurti, K. von
Klitzing, J. H. Smet, W. B. Johnson, and V. Umansky, Phys. Rev. B
\textbf{69}, 161306 (2004); Phys. Rev. B \textbf{70}, 155310
(2004); R. G. Mani, Physica E (Amsterdam) \textbf{22}, 1 (2004);
\textit{Advances in Solid State Physics}, Vol. \textbf{44}, edited
by B. Kramer (Springer, Heidelberg, 2004) p. 135; Physica E
\textbf{25}, 189 (2004); Appl. Phys. Lett. \textbf{85}, 4962
(2004); IEEE Trans. on Nanotech. \textbf{4}, 27 (2005); Int. J.
Mod. Phys. B \textbf{18}, 3473 (2004); Microelectron. J.
\textbf{36}, 366 (2005).

\bibitem{2} M. A. Zudov, R. R. Du, L. N. Pfeiffer, and K. W. West, Phys. Rev. Lett.
\textbf{90}, 046807 (2003).

\bibitem{3} S. I. Dorozhkin, JETP Lett. \textbf{77}, 577 (2003).

\bibitem{4} S. A. Studenikin, M. Potemski, P.T. Coleridge, A. Sachrajda, and Z.R. Wasilewski, Sol. St. Comm. \textbf{129}, 341
(2004).

\bibitem{5} R. L. Willett, L. N. Pfeiffer, and K. W. West, Bull.
Am. Phys. Soc. \textbf{48}, 459 (2003).

\bibitem{6} A. E. Kovalev, S. A. Zvyagin, C. B. Bowers, J. L. Reno, and J. A. Simmons, Sol. St. Comm. \textbf{130}, 379
(2004).

\bibitem{7} V. I. Ryzhii,  Sov. Phys. - Sol. St. \textbf{11}, 2078
(1970); V. Ryzhii and R. Suris, J. Phys. Conden. Mat. \textbf{15},
6855 (2003); V. Ryzhii and A. Satou,  J. Phys. Soc. Jpn.
\textbf{72}, 2718 (2003).

\bibitem{8} J. C. Phillips,  Sol. St. Comm. \textbf{127}, 233
(2003).

\bibitem{9} A. C. Durst, S. Sachdev, N. Read, and S. M. Girvin, Phys. Rev.
Lett. \textbf{91}, 086803 (2003); A. C. Durst and S. M. Girvin,
Science \textbf{304}, 1752 (2004).

\bibitem{10} A. V. Andreev, I. L. Aleiner, and A. J. Millis, Phys. Rev.
Lett. \textbf{91}, 056803 (2003).

\bibitem{11} J. Shi and X. C. Xie, Phys. Rev. Lett. \textbf{91}, 086801
(2003).

\bibitem{12} X. L. Lei and S. Y. Liu, Phys. Rev. Lett. \textbf{91}, 226805
(2003); X. L. Lei, J. Phys.: Condens. Matter \textbf{16}, 4045 (2004).

\bibitem{13} I. A. Dmitriev, A. D. Mirlin, and D. G. Polyakov, Phys. Rev.
Lett. \textbf{91}, 226802 (2003).

\bibitem{14} K. Park, Phys. Rev. B \textbf{69}, 201304 (2004).

\bibitem{15} M. G. Vavilov, I. A. Dmitriev, I. L. Aleiner, A. D. Mirlin, and
D. G. Polyakov, Phys. Rev. B \textbf{70}, 161306 (2004).

\bibitem{16} C. Joas, M. E. Raikh, and F. von Oppen, Phys. Rev. B \textbf{70}, 235302
(2004).

\bibitem{17} J. Inarrea and G. Platero, Phys. Rev. Lett. \textbf{94}, 016806
(2005).

\bibitem{18} F. S. Bergeret, B. Huckestein, and A. F. Volkov, Phys. Rev. B
\textbf{67}, 241303 (2003); A. A. Koulakov and M. E. Raikh, Phys.
Rev. B \textbf{68}, 115324 (2003); P. H. Rivera and P. A. Schulz,
Phys. Rev. B \textbf{70}, 075314 (2004); S. A. Mikhailov, Phys.
Rev. B \textbf{70}, 165311 (2004); T. Toyoda, M. Fujita, H.
Koizumi, and C. Zhang, Phys. Rev. B \textbf{71}, 033312, (2005).
A. E. Patrakov and I. I. Lyapilin, Low. Temp. Phys. \textbf{30},
874 (2004); J. Dietel, L. I. Glazman, F. W. J. Hekking, and F. von
Oppen, Phys. Rev. B \textbf{71}, 045329 (2005); M. Torres and A.
Kunold, Phys. Rev. B \textbf{71}, 115313 (2005); Phys. Status Solidi
B \textbf{242}, 1192 (2005); M. Kennett, J. P.
Robinson, N. R. Cooper, and V. I. Falko, Phys. Rev. B \textbf{71},
195420 (2005); M. Oswald and J. Oswald, Int. J. Mod. Phys. B
\textbf{18}, 3489 (2004); M. Oswald, J. Oswald, and R. G. Mani,
Phys. Rev. B \textbf{72}, 036334 (2005); A. Auerbach, I. Finkler, B. I.
Halperin, and A. Yacoby, Phys. Rev. Lett. \textbf{94}, 196801
(2005); Yu. V. Pershin and C. Piermarocchi, Appl. Phys. Lett. 86,
212017 (2005); K. Ahn, cond-mat/0504228.

\bibitem{19} F. F. Fang and P. J. Stiles, Phys. Rev. \textbf{174}, 823
(1968); T. Ando, A. B. Fowler, and F. Stern, Rev. Mod. Phys.
\textbf{54}, 1 (1982).

\bibitem{20} C. L. Yang, R. R. Du, L. N. Pfeiffer, and K. W. West,
cond-mat/0504715.

\bibitem{21} For this discussion, the characteristic field scale of the 2D
Shubnikov-de Haas effect in the GaAs/AlGaAs system is the magnetic
field that helps to realize the filling factor $\nu$ = 1
condition, when the sample is oriented perpendicular to the
magnetic field. As the sample is tilted by $\theta$ with respect
to the applied $B$, the SdH oscillations span a larger $B$ scale
proportional to $cos^{-1}(\theta)$, indicating that the SdH
characteristic field scale is determined by $B_{\perp}$.

\end{thebibliography}
\end{document}